\documentclass[3p,times]{elsarticle}

\usepackage{ecrc}

\volume{00}

\firstpage{1}

\journalname{Nuclear Physics A}

\runauth{Olena Romanets et al.}

\jid{NUPHA}

\jnltitlelogo{Nuclear Physics A}

\CopyrightLine{2012}{Published by Elsevier Ltd.}

\usepackage{amssymb}

\newcommand{\SU}{{\rm SU}}

\newcommand{\MeV}{{\rm MeV}}

\newcommand{\ignore}[1]{} 

\usepackage[figuresright]{rotating}

\begin{document}

\begin{frontmatter}

\dochead{}



\title{Heavy-quark spin symmetry for charmed and strange baryon resonances}

\author[a]{Olena~Romanets}\author[b,c]{Laura~Tolos}\author[d]{Carmen~Garc\'ia-Recio}\author[e]{Juan~Nieves}\author[d]{Lorenzo~Luis~Salcedo}
\author[a]{Rob~Timmermans}

\address[a]{Theory Group,  KVI, University of Groningen, Zernikelaan~25, 9747~AA~Groningen, The~Netherlands}

\address[b]{Instituto de Ciencias del Espacio (IEEC/CSIC), Campus Universitat Aut\`onoma de Barcelona, Facultat de Ci\`encies, Torre C5, 
E-08193 Bellaterra~(Barcelona), Spain}

\address[c]{Frankfurt Institute for Advanced Studies, Johann Wolfgang Goethe University, Ruth-Moufang-Str. 1, 60438 Frankfurt am Main, Germany}

\address[d]{Departamento~de~F\'isica~At\'omica, Molecular~y~Nuclear, and Instituto Carlos I de F{\'\i}sica Te\'orica y Computacional, Universidad~de~Granada,
 E-18071~Granada, Spain}

\address[e]{Instituto~de~F\'isica~Corpuscular~(centro~mixto~CSIC-UV), Institutos~de~Investigaci\'on~de~Paterna, Aptdo.~22085,~46071,~Valencia,  Spain}

\begin{abstract}
 We study charmed and strange odd-parity baryon resonances that are generated dynamically by a unitary
 baryon-meson coupled-channels model which incorporates heavy-quark spin symmetry. This is
 accomplished by extending the SU(3) Weinberg-Tomozawa chiral Lagrangian to SU(8) spin-flavor 
symmetry plus a suitable symmetry breaking. The model generates resonances with
 negative parity from the s-wave interaction of pseudoscalar and vector mesons with $1/2^+$ and $3/2^+$
 baryons in all the isospin, spin, and strange sectors with one, two,
 and three charm units.
 Some of our results can be identified with experimental data 
 from several facilities, such as the CLEO, Belle, or BaBar Collaborations, as well as with other
 theoretical models, whereas others do not have a straightforward identification and require the
 compilation of more data and also a refinement of the model.
\end{abstract}


\end{frontmatter}


\section{Introduction}
\label{introduction}
%
%
%

The properties of heavy-flavored hadronic resonances have attracted a lot of attention lately. 
The study of such states can help in the interpretation of the nature of particles found in past and 
ongoing experiments (e.g. CLEO, BaBar, Belle, LHCb)
~\cite{experiments},
as well as in understanding states which will be 
discovered in future experiments (e.g. PANDA at GSI~\cite{PANDA}). 
It is important to understand whether baryon 
(meson) resonances can be described as excited three-quark (quark-antiquark) states or rather as 
hadron molecules; also a combined interpretation of such states is possible. 

At present
 there is a lack of a robust scheme to systematically construct an effective field theory approach to
 study four flavor physics. 
Some steps in that direction have been taken  by recent studies using coupled-channels models. Among them one can find unitarized coupled-channels
models
\cite{unitarizedcchm, TVME, JimenezTejero:2009vq, Hofmann:2006qx},
the J\"ulich meson-exchange model~\cite{Julich}
and schemes based on hidden gauge formalism~\cite{Wu:2010jy}.
These models are not fully consistent with the heavy-quark spin
symmetry (HQSS)
\cite{HQSS},
which is a proper symmetry of Quantum Chromodynamics  (QCD) in the limit of infinitely heavy quark masses.
There have also been some attempts to build a scheme based on chiral perturbation theory for hadrons 
which contain heavy quarks
\cite{WiseCasalbuoniYan}. Moreover, an effective theory which incorporates heavy-quark, chiral  and 
hidden local gauge symmetries was developed for studying baryon-baryon interactions 
\cite{Oka}.
Besides, an SU(8) spin-flavor symmetric unitarized coupled-channels model has been recently 
developed and used 
for various numbers of charm and strangeness
\cite{SU(8), Gamermann}.

In this paper we study dynamically-generated baryon resonances, using the SU(8) spin-flavor model. We have payed special attention to analyze the underlying symmetry of the interaction. In particular, we
have studied the original group multiplets from where each of the found baryon resonances originates
and obtained the different HQSS multiplets.
Our studies covered states with charm and strangeness,
and in the following sections we will show and discuss our results 
for baryon resonances with charm $C=1$, namely
 $\Lambda_c$  (strangeness $S=0$, isospin $I=0$), 
$\Sigma_c$ ($S=0$, $I=1$), 
$\Xi_c$ ($S=-1$, $I=1/2$)
and $\Omega_c$ ($S=-2$, $I=0$).

\section{Theoretical framework}
We use the SU(8) spin-flavor model of \cite{SU(8), Gamermann}.
The interaction potential is an extension of the SU(3) chiral Weinberg-Tomozawa
potential to the SU(8) symmetry.
In this model vector mesons are treated on equal footing with pseudoscalar mesons
and both spin-1/2 and spin-3/2 baryons are taken into account. 
We only consider s-wave interaction, which is appropriate close to the meson-baryon thresholds.
In this SU(8) scheme the mesons fall in the {\bf 63}-plet, and the baryons are placed in
the {\bf 120}-plet. Consequently, in the s-channel, the baryon-meson space reduces into 
four SU(8) irreps, three of which ({\bf 120}, {\bf 168} and {\bf 4752}) are attractive. 
We find that the multiplets {\bf 120} and {\bf 168}
are the most attractive ones, and
therefore we have 
concentrated our study on the states which belong to these two irreps
in the SU(8) symmetric limit.

We consider the reduction of the SU(8) symmetry 
${\rm SU(8)} \supset {\rm SU(6)}\times {\rm SU}_C(2) \times {\rm U}_C(1)$, 
where SU(6) is the spin-flavor group for three light flavors, ${\rm SU}_C(2)$ is 
the charm quark rotation group, and ${\rm U}_C(1)$ is the group generated by the 
charm quantum number $C$.
The SU(6) multiplets can be reduced under $\SU(3)\times \SU_l(2)$, where
$\SU_l(2)$ refers to the spin of the light quarks.
We further
reduce $\SU_l(2) \times \SU_C(2)\supset\SU(2)$ where $\SU(2)$ refers to the
total spin $J$;
in this way we make the connection with the labeling ($C, S, I, J$).

The contact tree-level meson-baryon interaction of the extended SU(8) symmetric
Weinberg-Tomozawa potential reads
\begin{equation}
V_{ij}(s)= D_{ij}
\frac{2\sqrt{s}-M_i-M_j}{4\,f_i f_j} \sqrt{\frac{E_i+M_i}{2M_i}}
\sqrt{\frac{E_j+M_j}{2M_j}} 
\,.
\label{eq:vsu8break}
\end{equation}
Here, $i$ and $j$ are the outgoing and incoming baryon-meson channels;
$M_i$, $E_i$, and $f_i$ stand, respectively, for the mass and the 
center-of-mass energy of the baryon and the meson decay constant in the $i$ channel;
 $D_{ij}$ are the matrix elements for the various
$CSIJ$ sectors considered in this work, which can be found in the Appendices 
of~\cite{SU(8), we}.

The scattering amplitudes
are calculated by solving the
on-shell Bethe-Salpeter equation in the coupled channels:
\begin{equation}
\label{LS}
 T(s) =\frac{1}{1-V(s)G(s)} V(s).
\end{equation}
Here $G(s)$ is a diagonal matrix containing the baryon-meson propagator for each
channel, and $D$, $T$ and $V$ are matrices in coupled-channels space.

The bare loop function $G^{0}_{ii}(s)$ is logarithmically ultraviolet
divergent and needs to be renormalized.
We have renormalized our amplitudes by using
a subtraction point regularization, with
a subtraction point $\sqrt{s}=\mu_i=\sqrt{m_{\rm th}^2+M_{\rm th}^2}$,
where $m_{\rm th}$ and $M_{\rm th}$ are, respectively, the masses of the meson and
the baryon of the channel with the lowest threshold in the given $CSI$ sector,
\begin{equation}
G_{ii}(s)= G^{0}_{ii}(s)-G^{0}_{ii}(\mu_i^2)
.
\label{eq:subs}
\end{equation}
In order to get a  better fine-tuning with the experimental data, one can define 
 the subtraction point as $\mu_i=\sqrt{\alpha~(m_{\rm th}^2+M_{\rm th}^2)}$,  with $\alpha$ being slightly different from one.

The baryon resonances are obtained as poles of the scattering amplitude on 
the complex-energy plane. 
\ignore{
The poles on the first Riemann sheet on the real energy axis below thresholds
are interpreted as bound states (though we usually call all found states baryon resonances),
the poles on the second Riemann sheet below the real axis and above thresholds are 
baryon resonances, the poles on the second Riemann sheet on or below the real axis 
but below thresholds are virtual states, and the other possible poles are considered as unphysical. 
}
The mass $m_R$ and the width $\Gamma_R$ of the state can be obtained from the coordinate $\sqrt{s_R}$ 
of the corresponding pole 
on the complex energy $\sqrt{s}$ plane, $\sqrt{s_R}=m_R-\frac{i}{2} \Gamma_R$,
and the couplings to the meson-baryon channels can be found from the residues 
of the $T$-matrix around the pole.
\ignore{
\begin{equation}
\label{Tfit}
 T_{ij} (s) \approx \frac{g_i g_j}{\sqrt{s}-\sqrt{s_R}}.
\end{equation}
}

The matrix elements $D_{ij}$ display exact SU(8) invariance, but this symmetry
is severely broken in nature. Therefore we implement symmetry-breaking
mechanisms. It should be mentioned here that we have removed the channels with 
the $c \bar c$
pairs to be consistent with the HQSS. 
In the present work
we use physical values for the masses of the hadrons and for the
decay constants of the mesons. 
The symmetry is broken following the chain 
$\SU(8)\supset\SU(6)\supset\SU(3)\supset\SU(2)$,
where the last group $\SU(2)$ refers to isospin.
This symmetry breaking was performed by adiabatic change of the
hadron masses and meson weak decay constants.
In this way we can label each baryon resonance with the
original group multiplet and define the HQSS multiplets.

\section{Dynamically generated baryon resonances}
Let us begin with the $\Lambda_c$ states.
Our model generates four $\Lambda_c$ baryon resonances,
three with spin $J=1/2$ and one with $J=3/2$.
By comparing the dominant channels with the 
decay channels of the experimental states,
two of our $\Lambda_c$'s 
have been
identified with experimentally known states.
We identify the experimental $\Lambda_c(2595)$ resonance with
the state that 
we found around $2618.8 - i 0.6\,\MeV$.
The experimental value of the width of $\Lambda_c(2595)$
$3.6^{+ 2.0}_{ -1.3}\,\MeV$ is not reproduced,
due to the fact that
we have not included the three-body decay
channel $\Lambda_c \pi \pi$, which already represents almost one third of the
decay events \cite{Nakamura:2010zzi}. 
Our result for
$\Lambda_c(2595)$ agrees with the results from $t-$channel vector-meson
exchange (TVME) models
\cite{TVME, JimenezTejero:2009vq}, but, as it was first pointed out in Ref.~\cite{SU(8)}, we claim a
dominant $N D^*$ component in its structure, 
whereas in the TVME model the $\Lambda_c(2595)$ is generated mostly 
as a
$N D$ bound state.
We also obtain a broad resonance with a mass very close to the 
 $\Lambda_c(2595)$,
 namely at
$2617.3\,\MeV$. It couples strongly to the open channel $\Sigma_c \pi$.
The other pole with $J=1/2$ that we find around
$2828 - i 0.4\,\MeV$
has not been identified with any known experimental state.

In the $J=1/2$ sector there are 16 coupled channels, which can generate $\Lambda_c$
resonances. Every found baryon resonance couples strongly
only to some of the coupled channels, see~\cite{we}. Therefore, we study how the features 
(masses, widths and couplings) of the $\Lambda_c$ resonances change, when
we consider only the dominant meson-baryon coupled channels.
It turns out that the masses and widths, as well as couplings do not change 
drastically when we only consider the restricted coupled-channels space.
The width of the $\Lambda_c(2617.3)$ resonance increases from 89.8 to 97.3~MeV,
whereas the mass
and the coupling to the $\Sigma_C \pi$ stay unchanged.
The $\Lambda_c(2618.8)$ resonance slightly increases its mass by 2.6~MeV, 
and the width
decreases from 1.2 to 1.1~MeV,
while the coupling to $N D^*$ channel
remains almost the same.
Finally, the mass of the $\Lambda_c(2828.4)$ state raises by 8.6~MeV,
and its width is now 1.0~MeV;
the couplings to the dominant $\Lambda_c \eta$  and $\Sigma_c^* \rho$ channels 
slightly vary.

Further, we find one $\Lambda_c$ resonance with $J=3/2$
 located at $(2666.6 -i 26.7\,\MeV)$.
 We identify this resonance with the experimental $\Lambda_c(2625)$
\cite{Nakamura:2010zzi}.
The experimental $\Lambda_c(2625)$ has a very narrow width, $\Gamma < 0.97\,\MeV$, and
decays mostly to $\Lambda_c \pi \pi$. 
By changing the subtraction point, such that the mass of the resonance
is closer to the value of the experimental one,
the phase space would be reduced. 
A similar resonance was found at $2660\,\MeV$ in the TVME model 
of Ref.~\cite{Hofmann:2006qx}. However, in our calculation
we obtain a non-negligible contribution from the baryon-vector meson
channels to the generation of this resonance, as already observed in
Ref.~\cite{SU(8)}.
When restricting the number of coupled channels to the four ones, to 
which $\Lambda_c^*(2666.6)$ couples the most, namely
$\Sigma_c^* \pi$, $N D^*$, $\Sigma_c \rho$ and $\Sigma_c^* \rho$, the resonance 
features are changed as follows. The mass somewhat increases by 1.2~MeV, while the width 
grows by 8.2~MeV, and couplings remain almost unchanged.

We obtain three spin-1/2 $\Sigma_c$ resonances,
with masses 2571.5, 2622.7 and 2643.4~MeV.
These states are predictions of our model, since there is no experimental data
in this energy region. 
In the SU(4) model of Ref.~\cite{JimenezTejero:2009vq}
two $\Sigma_c$ spin-1/2 resonances
are predicted.
In this reference, the first $\Sigma_c$ resonance has a mass 
$2551\,\MeV$ and a width of $0.15\,\MeV$, and
it can be associated with the
$\Sigma_c(2572)$ state of our model which we generate
with the width $\Gamma=0.8\,\MeV$. 
However, in our
model this resonance couples most strongly to the channels which
incorporate vector mesons, 
whereas in Ref.~\cite{JimenezTejero:2009vq} it is not the case.
The other resonance predicted in Ref.~\cite{JimenezTejero:2009vq} 
cannot be compared to any of
our results.
Further, we obtain two spin-$3/2$ $\Sigma_c$ resonances.
The first one, a bound state at 2568.4~MeV, 
lies below the threshold of any possible 
decay channel
and  is thought to be the charmed counterpart of the
hyperonic $\Sigma(1670)$ resonance. 
The second state at $2692.9 -i 33.5$ ~MeV has no direct comparison with the
available experimental data.

Our model generates six $\Xi_c$ states with $J=1/2$ and three ones 
with $J=3/2$.
In this sector there are two negative-parity experimentally known resonances
that can be identified with some of our dynamically-generated states, namely experimental
$\Xi_c(2790)~J^P=1/2^-$ and
$\Xi_c(2815)~J^P=3/2^-$ \cite{Nakamura:2010zzi}. 
The state $\Xi_c(2790)$ has a width of $\Gamma<12-15\,\MeV$ and it decays to
$\Xi_c' \pi$, with $\Xi_c' \rightarrow \Xi_c \gamma$. 
We assign it to the
$2804.8 -i 13.5$~MeV state found in our model because of the large 
$\Xi'_c\pi$ coupling.
A slight modification of the subtraction point can lower the
position of our resonance to $2790\,\MeV$ and most probably reduce its width as it will get
closer to the $\Xi_c' \pi$ channel.
\ignore{Moreover, this seems to be a reasonable
assumption in view of the fact that, in this manner, this $\Xi_c$ state is the
HQSS partner of the $\Xi_c^*(2845)$ state, which we will identify with the
$\Xi_c^*(2815)$ resonance of the PDG.}
It could be also possible to
identify our pole at $2733\,\MeV$ 
with the experimental $\Xi_c(2790)$ state. In that case, one
would expect that if the resonance position gets closer to the physical mass
of $2790\,\MeV$, its width will increase and it will easily reach values of
the order of $10\,\MeV$.
The full width of the experimental $\Xi_c$ resonance with $J^P=3/2^-$  is
expected to be less than $3.5\,\MeV$ for $\Xi_c^+(2815)$ and less than
$6.5\,\MeV$ for $\Xi_c^0(2815)$, and the decay modes are $\Xi_{c}^+ \pi^+
\pi^-$, $\Xi_{c}^0 \pi^+ \pi^-$. We obtain two resonances at $2819.7 -i 16.2\,\MeV$ and
$2845.2 -i 22.0\,\MeV$, respectively, that couple strongly to $\Xi_c^* \pi$, with
$\Xi_c^* \rightarrow \Xi_c \pi$. Allowing for this possible indirect
three-body decay channel, we might identify one of them to the experimental
result. This assignment is possible for the state at $2845.2\,\MeV$
if we slightly change the subtraction point, which will lower its
position and reduce its width as it gets closer to the threshold of the open $\Xi_c^* \pi$
channel.

We obtain three $\Omega_c$ bound states with masses $2810.9$, $2884.$5 and 
$2941.6$~MeV.
There is no experimental information on those excited states. However, our
predictions can be compared to recent calculations of
Ref.~\cite{JimenezTejero:2009vq}.
In this work three $\Omega_c$ resonances are predicted,
with masses higher than the ones of our resonances by approximately 100~MeV.
Further, we obtain two spin-$3/2$ bound  states $\Omega_c$
with masses $2814.3$ and
$2980.0$~MeV, which mainly couple to $\Xi D^*$ and $\Xi^* D^*$, and to $\Xi_c^* \bar K$,
respectively. As in the $J=1/2$ sector, no experimental information is
available here.

\section{Summary}
Charmed baryon resonances, in particular $\Lambda_c$, $\Sigma_c$, $\Xi_c$ 
and $\Omega_c$ odd-parity states 
have been studied within a coupled-channels unitary approach 
that implements HQSS.
For this purpose the SU(8) spin-flavor symmetric model of Ref.~\cite{SU(8)} has been used. We have obtained four
$\Lambda_c$ baryon resonances,  two of which can be identified with the
experimental $\Lambda_c(2595)$ and $\Lambda_c(2625)$ states.
When the number of coupled channels is reduced to the dominant ones, 
the features (mass, width, coupling constants)
of the corresponding resonance do not change significantly. 
Further, five $\Sigma_c$ and nine $\Xi_c$ resonances are obtained.
Some of our resonances can be identified with experimentally known $\Sigma_c$ and $\Xi_c$ states,
 while others require the compilation of more data and a refinement of the model.

\section{Acknowledgments}
 This research was supported by
DGI and FEDER funds, under contracts FIS2011-28853-C02-02, FIS2011-24149,
FPA2010-16963 and the Spanish Consolider-Ingenio 2010 Programme CPAN
(CSD2007-00042), by Junta de Andaluc{\'\i}a grant FQM-225, by Generalitat
Valenciana under contract PROMETEO/2009/0090 and by the EU HadronPhysics2
project, grant agreement n. 227431. O.R. wishes to acknowledge support from
the Rosalind Franklin Fellowship. L.T. acknowledges support from Ramon y Cajal
Research Programme, and from FP7-PEOPLE-2011-CIG under contract
PCIG09-GA-2011-291679.








\end{document}